# A Testing Time for Anti-matter

*Precision measurement of anti-protonic Helium provides a test of physical laws*

Wim Ubachs

Department of Physics, Vrije Universiteit, De Boelelaan 1081, 1081 HV Amsterdam, The Netherlands.
Email: w.m.g.ubachs@vu.nl

Spectroscopy is the most accurate branch of science. Optical transition frequencies in isolated atoms and molecules can nowadays be measured to many-digit accuracies applying the tools developed in the atomic physics community: ultra-stable lasers, locked via frequency-comb lasers to atomic clocks, and the techniques to cool and control the motion of atoms. Precision measurements on small quantum systems can be compared with its theoretical descriptions at the most fundamental level to subject physics theories to a test and to search for physics beyond the Standard Model of Physics [1]. On page 610 of this issue Hori et al. [2] apply these tricks of the trade to a small atomic quantum system with a built-in anti-particle to perform very precise spectroscopic measurement in anti-protonic helium (see Fig. 1). For the first time the technique of buffer gas cooling is demonstrated on a composite matter-anti-matter particle therewith beating the Doppler effect in the experiment.

The laws of physics conspire to create the conditions for producing relatively long-lived states in the exotic anti-protonic helium atom (e$\bar{p}$He). In all other cases of anti-protons stopping in matter, the characteristic annihilation process by which a matter particle and an anti-matter particle end their lives in a flash of energy, takes places on a timescale below a nanosecond. The Heisenberg uncertainty-principle then prohibits a precision measurement. Capture of anti-protons in a helium atom, under replacement of an electron, is the exception [3]. The inclusion of the anti-proton produces a $\bar{p}$He$^+$ ionic core in a heavy Rydberg two-particle state. Energy conservation dictates that $\bar{p}$He$^+$ emerges in a quantum state of typically $n$=38; in the experiment by Hori et al. states with $n$=31 to $n$=40 are probed. The overlap of the wave functions of $\bar{p}$ and He$^{2+}$ is minimal if the high-$n$ Rydberg states exhibit a large angular momentum, for example in a state like $n$=38 with $l$=36 or 37. These circular states have typical lifetimes of microseconds, long enough to perform a laser experiment: narrow spectral lines can be measured to determine level energies to high accuracy. The Rydberg constant of the heavy Rydberg $\bar{p}$He$^+$ ion-pair is so large that the transitions by which the $n$-quantum number changes by one unit, lie in the wavelength range of the near-infrared, visible and ultraviolet, where Hori and co-workers have narrowband and stable lasers at hand. This is another fortunate circumstance.

The second electron of the helium atom remains in its 1s-orbital, and it acts as a spectator during the capture process. It favorably shields the composite newly-built electrically neutral e$\bar{p}$He-system during collisions with the outside world. This property makes it possible that the anti-protonic helium atom survives, while undergoing multiple collisions in the surrounding bath of cold helium atoms at a temperature of 1.5 degrees above absolute zero. This buffer gas cooling technique, developed for producing cold atoms and molecules [4], is applied for the first time to an exotic atom bearing an anti-particle in its structure. After a number of collisions, without annihilation, the exotic atoms are equilibrated and take up the temperature of the bath. At 1.5 Kelvin the velocity of the particles is very low, and hence the Doppler effect is evaded, allowing to measure narrow spectral lines.

The precision frequency measurements on 13 transitions in anti-protonic helium, performed at 9-digit accuracy, can be interpreted in several ways. First of all it is a test of CPT invariance, arguably the most fundamental theorem on symmetries in physics. All calculations in particle physics should loose validity if CPT symmetry is not valid [3]. Exchange of particles by their anti-particles (C), reflection in a mirror (P), and letting time run backwards (T) should not cause a change in the measurable properties of a physical system. The measurements verify that CPT is a valid symmetry, at least at the level of accuracy now obtained.

Anti-protonic helium is a three-particle Coulomb system, the stability of which was already investigated by Poincaré in the framework of classical mechanics. In the fully quantized version of electromagnetism (QED) the stability of the Helium atom [7] and the $H_2^+$ molecular ion [5] is well established and the bound state level energies can be calculated to 10-digit accuracy. The same is true for the ep̄He system [5,6] that is found to withstand a test of QED.

The transition frequencies sensitively depend on the mass ratios of the constituent particles. Similarly as in a spectroscopic determination of the proton-electron mass ratio from $HD^+$ spectroscopy [8] evaluation of Hori's data yields a value for the anti-proton-electron mass ratio. We learn that anti-protons weigh the same as protons, up to the 10$^{th}$ digit.

The frontier of particle physics is commonly approached in the high energy sector, employing accelerators like the Large Hadron Collider. The detection of the Higgs boson at the LHC earmarks the culmination of the Standard Model of Physics, but the quest is out to explore new physics beyond the celebrated Standard Model. Alternatives approaches exist in the low energy domain by performing extreme precision measurements on small atomic and molecular systems for which the quantum level structure is calculable [1]. This can be done through the search for an electric dipole moment of the electron in molecules that might reveal signatures of supersymmetry [9]. Laser precision measurements of optical transitions in molecules constrain the existence of higher dimensions beyond the 3+1 that we regularly observe, or the occurrence of fifth forces beyond the three forces known in the Standard Model plus gravity [1]. Previous measurements on anti-protonic helium have already shown to limit the strength of such hypothetical fifth force at the sub-Ångström length scale [10], and the present data constrain these phenomena even further. This exotic atom involving anti-matter seems a fortunate accident of nature. It exhibits long-lived (metastable) quantum states that can be probed with lasers, and it survives the collisional conditions needed to cool its kinetic motion, as is demonstrated now. These properties may be further exploited to reveal new physics in future experiments on this extraordinary atom-like particle.

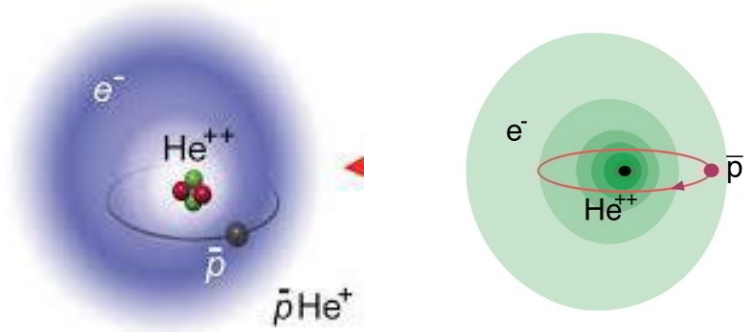

Fig. 1: Structure of the anti-protonic helium system (e$\bar{p}$He). In a (crude) approximation the positive He$^{++}$ and negative $\bar{p}$ ions can be considered to be bound as a heavy Rydberg system with Kepler radius of R(n)=$a_0 n^2/MZ^2$, with $a_0$ the Bohr radius, Z=2, and M the scaling factor for heavy Rydberg states [11], M=$(M_{\bar{p}} M_{He})/(M_{\bar{p}}+M_{He})m_e$=1224. For n=38, the typical state in which the anti-proton is captured, this yields a characteristic distance of 0.16 Å. In a further approximation the electron is then bound in the field of the two-particle ionic core of charge 1, similarly as in the hydrogen atom at a typical distance of the Bohr radius $a_0$ = 0.5 Å. The quantum level structure of anti-protonic helium can be calculated to 10-digit precision [5,6].